\documentclass[journal]{IEEEtran}

\usepackage{graphicx}
\usepackage{booktabs}
\usepackage{amsmath,amssymb}
\usepackage{fixltx2e}
\usepackage{float}
\usepackage{comment}
\usepackage{color} 
\usepackage{dblfloatfix}
\usepackage{float}
\usepackage{gensymb}
\usepackage{algorithm,algorithmic}
\usepackage{cite}

\newcommand{\bldW}{\mathbf W}
\newcommand{\bldDlt}{\boldsymbol \Delta}
\newcommand{\bldzero}{\boldsymbol 0}
\DeclareMathOperator{\var}{var}

\begin{document}

\title{Wireless Picosecond Time Synchronization for Distributed Antenna Arrays with Dynamic Connectivity}
\author{Naim Shandi,~\IEEEmembership{Graduate~Student~Member,~IEEE,} Jason M.\ Merlo,~\IEEEmembership{Graduate~Student~Member,~IEEE,} \\and Jeffrey A.\ Nanzer,~\IEEEmembership{Senior Member,~IEEE}
\thanks{Manuscript received 2024.}
\thanks{This work was supported in part by Office of Naval Research under Grant \#N00014-20-1-2389, the National Science Foundation under Grant \#1751655, and Google under the Google Research Scholar program \textit{(Corresponding author: Jeffrey A.\ Nanzer)}}
	\thanks{	
	The authors are with the Department of Electrical and Computer Engineering, Michigan State University, East Lansing, MI 48824 USA (email: shandina@msu.edu, merlojas@msu.edu, nanzer@msu.edu).}
}

\maketitle

\begin{abstract}
Phase, time, and frequency coordination are crucial for the coherent operation of distributed antenna arrays. This paper demonstrates a high accuracy decentralized time synchronization method for arrays with dynamic connectivity. To overcome challenges such as alteration in network topology, system scaling, and single- or multiple-node failures, we present a decentralized time synchronization method implementing the average consensus algorithm and two-way time transfer. Our approach supports decentralized time synchronization with picosecond accuracy ($< 13$ ps), unbiased convergence, resilience to node failures, and robustness to other network connection adversities. A time synchronization precision of under 3 ps was achieved for a fully connected frequency syntonized network. The algorithm ensured convergence in time even for a network with only one link connecting two randomly chosen nodes at each iteration; however, the speed of convergence decreased along with the number of connected links. The method is exhibited for a six-node distributed array, by both simulation and experimental implementation using software defined radios.
\end{abstract} 


\section{Introduction}

Wireless coordination of distributed antenna arrays has increased in significance for emerging networked and distributed communication and sensing systems. Applications include remote sensing, distributed beamforming, navigation, and global positioning \cite{nasa2015roadmap,4202181}. Compared to single-platform systems, coherently operated distributed antenna arrays offer cost-effective increases in signal gain, robustness to element failures, and dynamic reallocation of array elements. However, realizing these advantages in applications such as high-gain beamforming requires accurate coordination of time, phase, and frequency across the array elements \cite{nanzer2021distributed}. Synchronization of clock times in a distributed system is especially challenging as it requires sub-nanosecond accuracy in time alignment. For coherent operation, the relative timing must be aligned to within a small fraction of the information duration of the signal, which may be on the order of tens of picoseconds for high bandwidth signals~\cite{8378649}. 

Previous work on distributed antenna array time synchronization focused mainly on centralized approaches relying on a reference node \cite{merlo2022wireless,9173801}. The vulnerability of such approaches to single point failure can be addressed through decentralized topologies where timing information is exchanged between adjacent nodes to coordinate functions and achieve consensus without a reference node~\cite{140489}.
Widely used for distributed coordination in multi-node networks is the consensus algorithm, through which all nodes can agree on a final decision or a parameter (e.g, time, phase, frequency) via an iterative process. When the agreed upon value is the average of the initial nodal states, the algorithm is termed average consensus \cite{5411807}; it appears in applications such as distributed computation \cite{10.5555/2821576} and distributed beamforming \cite{ouassal2021decentralized}. Some researchers focused on consensus within fixed topologies where the number of nodes and their connections stay fixed or stationary \cite{xiao2004fast}; others studied average consensus in switching or dynamic network topologies where the number of nodes and their connections change during the consensus process \cite{1657268}. Under ideal conditions (noiseless communication channels and strongly connected networks), average consensus always converges to the average of the initial nodal states \cite{xiao2004fast}. However, practical internode communication is corrupted by factors such as channel noise, thermal noise, fading, and quantization \cite{5411807}.
Average consensus has produced high accuracy distributed coordination (phase alignment, time synchronization, and frequency syntonization) in wireless sensor networks and distributed antenna arrays~\cite{ouassal2021decentralized,shandi2023}. In~\cite{shandi2023} the authors implemented decentralized time synchronization using a modified average consensus algorithm \cite{ouassal2021decentralized} combined with a high-accuracy two-way time transfer technique \cite{merlo2022wireless} to achieve picosecond accuracy for a fixed-topology distributed antenna array. 

The present work applies the decentralized wireless time synchronization method proposed in \cite{shandi2023} to a distributed antenna array with a dynamic connectivity graph where the number of internode connections changes randomly over time. In a frequency locked system, there still are time-varying parameters (e.g., temperature-related group delays) which manifest as a slow time rate change of phase (i.e., frequency offset) this algorithm can compensate for. Performance of the method is evaluated, through both simulation and experiment, for a multi-node array having dynamic network connectivity. The system achieves a decentralized high-accuracy wireless time synchronization error of less than ten picoseconds and a precision of under three picoseconds. Performance evaluation under the studied conditions shows that the system achieves and maintains time synchronization under conditions of dynamic connectivity with varying convergence rates even when just one internode link is randomly cycling around the array. These results provide a basis for reliable, high-accuracy time synchronization for dynamic distributed antenna arrays and networked systems in general.

\section{Decentralized Time Alignment}
\subsection{Average Consensus Method}
The algorithm evaluated in this paper is based on the average consensus algorithm~\cite{shandi2023, 140489} combined with the two-way time transfer method~\cite{merlo2022wireless,Levine_2008} to achieve picosecond-level wireless time synchronization for the nodes in a decentralized topology. The distributed antenna array is modeled as a graph $G=\{N,\mathcal{E}\}$ where $N=\{1,2, \ldots, n\}$ is a set of $n$ nodes and $\mathcal{E}$ is a set of undirected edges (i.e., communication is bidirectional along each edge; the notation $(i,j) \in \mathcal{E}$ means that the pair of nodes $i$ and $j$ is connected \cite{ouassal2021decentralized}). The local clock in each node is expressed as a function of the true global time $t$. For node $i$ in a frequency locked network, the local time is given by
\begin{equation}\label{clock}
    T_i(t)=t+\alpha_i+\beta_i(t)+\nu_i(t)
\end{equation}
where $\alpha_i$ is the static component of the time-varying bias or the clock time offset, and $\beta_i(t)$ is the dynamic component of time-varying bias accounting for the effects of temperature-dependent group-delays. The noise function $\nu_i(t)$ models the effects of zero-mean noise sources such as thermal, shot, flicker, and quantum noise types \cite{pozar2005microwave}.
The main assumption is that any time-varying frequency offset is quasistatic over the synchronization epoch \cite{merlo2022wireless}; its validity is established by locking the frequency to ensure that the nodes are syntonized, which can be accomplished in various ways such as two-tone frequency locking syntonization \cite{9223809}.
In this work we correct for both time bias components, $\alpha_i$ and $\beta_i(t)$, to achieve high accuracy time synchronization by compensating for the biases at the RF front-ends which would cause drifts over time even when the clocks are frequency locked. 
Let $\Delta_{ji}$ be the relative time offset at the RF front-ends between node $i$ and any adjacent node $j$, where a node is described as adjacent if $(i,j) \in \mathcal{E}$. These offset values are determined via the two-way time transfer approach, which relies on exchange of delay estimation waveforms between pairs of nodes. Four timestamps $T_j(t_{{\rm RX}j})$, $T_i(t_{{\rm TX}i})$, $T_i(t_{{\rm RX}i})$, and $T_j(t_{{\rm TX}j})$ are used to calculate $\Delta_{ji}$ for $j \ne i$ as 
\begin{equation}\label{delay_estimate}
\Delta_{ji} = \frac{[T_j(t_{{\rm RX}j}) - T_i(t_{{\rm TX}i})]-[T_i(t_{{\rm RX}i}) - T_j(t_{{\rm TX}j})]}{2},
\end{equation}
where $t_{RX_i}$ and $t_{TX_i}$ denote the true receiving and transmitting times at node $i$, respectively. Note that $\Delta_{ii} = 0$ for all $i$. After the $\Delta_{ji}$ are estimated, average consensus produces decentralized time synchronization by correcting for the average time offset $(\bldW\bldDlt)_{ii}$ at node $i$ which is estimated from all nodes at iteration $k$ with respect to the adjacent connected nodes. Here $\bldDlt = [\Delta_{ji}]$, and $\bldW = [w_{ji}]$ is the $n \times n$ real mixing matrix (created using the Metropolis--Hastings constant edge weight matrix) which has nonzero entries corresponding to the edges in the graph and self loops (diagonal entries). The properties of $\bldW$ required to ensure convergence are discussed in~\cite{boyd}. 

Average consensus requires only local time information sharing between a node and its neighbors in a strongly connected network. In such a network with static topology and mixing matrix satisfying needed constraints \cite{boyd}, the clock will converge to the global average. The dynamic time errors of a realistic system are addressed by using a synchronization interval sufficiently shorter than clock drift in addition to the continuous frequency syntonization of the system. During each iteration, the clock times at all connected nodes are updated using the decentralized time synchronization equation
\begin{equation}\label{DTSE}
   t_i(k)=t_i(k-1)+(\bldW\bldDlt)_{ii}(k-1) ,
\end{equation}
where $k$ is the iteration number, $t_i(k)$ is the updated time at node $i$, and $t_i(k-1)$ and $(\bldW\bldDlt)_{ii}(k-1)$ are the previous clock time and average time offsets, respectively, at node $i$. The system achieves convergence when the clock time at each node reaches the average of the initial clock times at all nodes (i.e., time offsets $\bldDlt\approx \bldzero$).

\begin{figure}[t!]
\centering
\includegraphics[width=0.65\columnwidth]{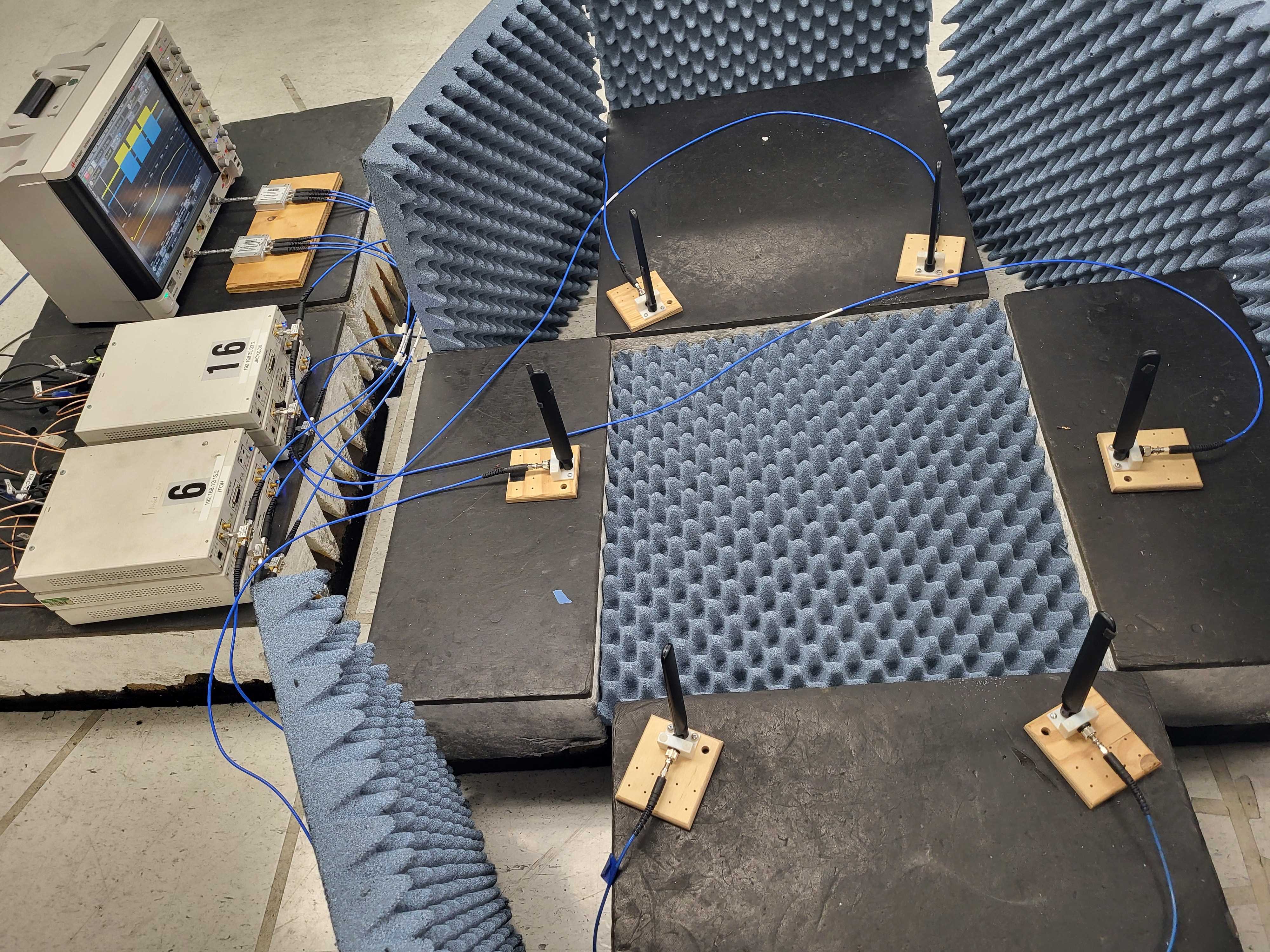}
\caption{Experimental setup. Six SDRs simulated the six-node array. Nodes 1 through 5 were daisy-chained to the 10 MHz frequency reference output and PPS output of node 0. Dipole antennas were used for wireless two-way time transfer between the nodes. RF absorber walls reduced multipath effects on the transmitted signal. TDMA was employed to schedule the time transfer between the nodes so that only one node could transmit at a time.}
\label{setup}
\end{figure}

\begin{figure*}[t!]
\centering
\includegraphics[width=1.3\columnwidth]{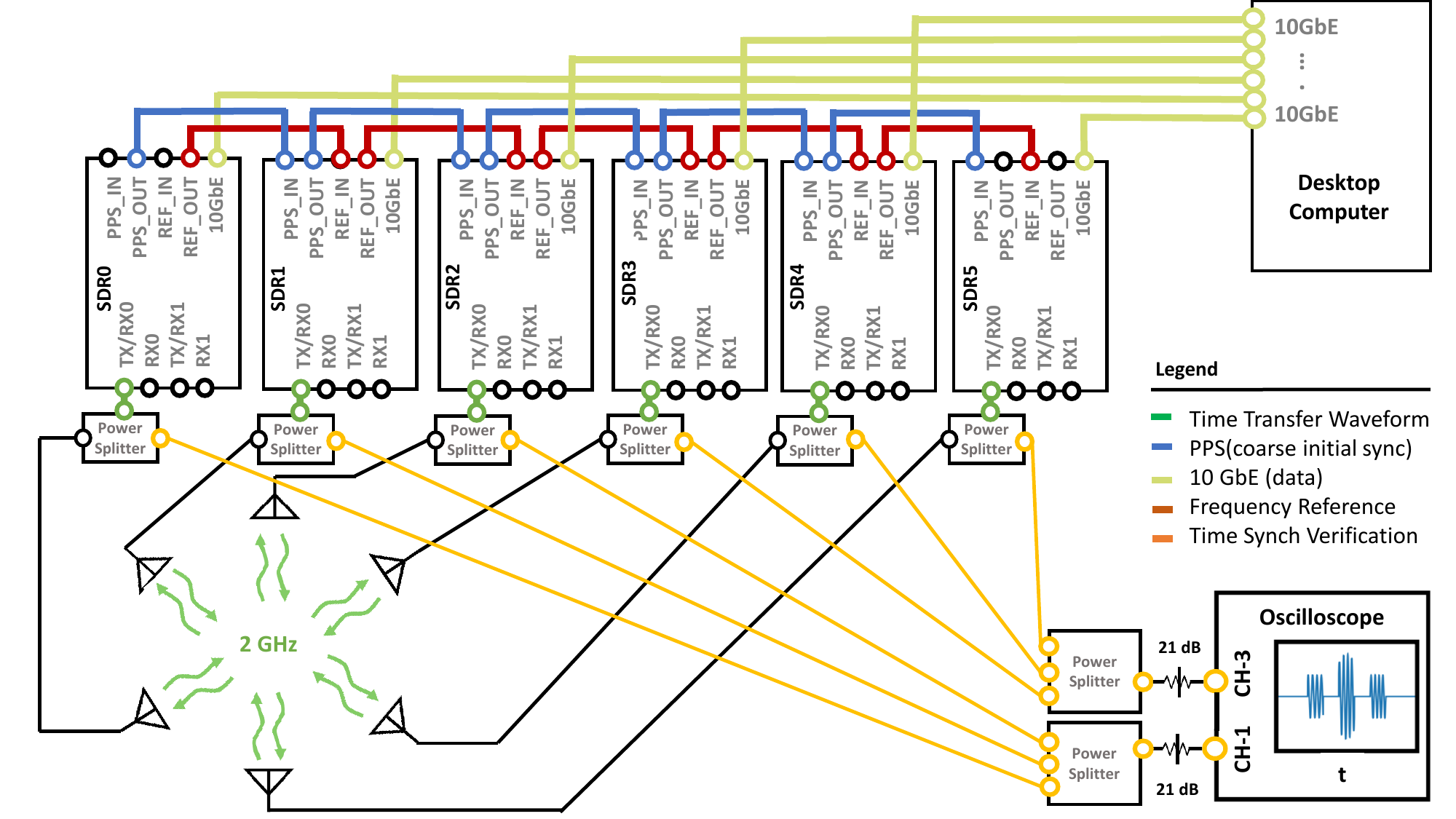}
\caption{System schematic. Six SDRs, controlled by one computer through 10 GbE cables, were scheduled to send two-tone waveforms for time offset and delay estimation. Each node exchanged waveforms with all others to estimate the relative clock offsets. Node 0 provided the initial coarse time alignment (within tens of nanoseconds) to the other nodes once at the start of the experiment. Connectivity between nodes was changed using the software.}
\label{schematic}
\end{figure*}

\subsection{Time Transfer Method}

The two-way time transfer method is based on the assumption of equal times of flight for messages traveling in opposite directions between two given nodes \cite{Levine_2008}. With this symmetry assumption, and its consequence that the channel is quasi-static during the synchronization epoch, two-way time transfer can be used to estimate both the time of flight and the time offsets between the RF front-ends of all nodes in the array \cite{merlo2022wireless}. This method has been used in applications including satellite clock synchronization \cite{68861} and network time synchronization protocols such as the precision time protocol (PTP) \cite{4579760}. The equations needed to estimate the time offset and propagation delay were given above. The primary source of error in the estimation of internode time offsets lies in the estimation of the receive timestamps. Hence the accuracy of the time delay estimator is crucial in reducing the estimation error. The theoretical bound for time delay estimation is given by the Cram\'er--Rao lower bound (CRLB) inequality $\var(\hat\tau-\tau) \geq 1/(2\zeta^2 \, \text{SNR})$. 
The left side is the variance of the delay estimate for a single link in the network. The right side contains the post-processed signal-to-noise ratio $\text{SNR} = E_s/N_0$ and the mean-square signal bandwidth $\zeta^2$ (i.e., the second moment of the signal spectrum) \cite{7801084}, \cite[Chapter 7.2]{richards2005fundamentals}. The error is minimized for a given SNR when $\zeta^2$ is maximized. This can be achieved when the signal energy is concentrated toward the edges of the available band (i.e., a two-tone signal) \cite{7801084}. The advantages of the approach, as well as the peak refinement method, are discussed in \cite{merlo2022wireless}. For a multinode network the CRLB limit becomes the average of the CRLB values for all $D$ connected links in the network \cite{shandi2023}:
\begin{equation}\label{CRLB2}
\var_{\text{ave}}(\hat\tau-\tau)\geq \frac{1}{2\zeta^2 D}\sum_{d=1}^D\frac{1}{\text{SNR}_d}.
\end{equation}

\section{Experimental Evaluation}
The performance of the decentralized time synchronization algorithm is evaluated in this section through simulations and experiments. The algorithm is evaluated through statistical analysis of the time synchronization results by calculating the average bias and the average standard deviation (precision).
Multiple laboratory experiments were conducted using software defined radios (SDRs) as nodes. The system was composed of six nodes (numbered 0 through 5) separated by approximately 45 cm (Fig.~\ref{setup}). Each node consisted of one Ettus Research X310 SDR with a UBX-160 daughterboard and a multiband swivel-mount dipole antenna (SPDA24700/2700). The SDRs were connected to a desktop computer through a 10 GbE cable and controlled by GNU Radio software. The TX/RX RF port of each SDR was connected to a one-to-two power splitter feeding both a dipole antenna through a six foot coax cable and a Keysight DSOS804A 20 GSa/s oscilloscope configured with an 8.4\;GHz analog bandwidth. The dipole was used for wireless time synchronization, while the oscilloscope captured the two-tone pulses which were used to evaluate the time alignment. Three-to-one power splitters were connected to channels 1 and 3 of the oscilloscope in order to receive the verification pulses from the two groups of three SDRs (Fig.\ \ref{schematic}). The SDRs were configured to transmit a two-tone waveform at carrier frequency of 2~GHz, bandwidth of 40~MHz, and pulse duration of 10 \textmu s with sampling rate of 200 MSa/s for wireless time synchronization purposes. They were also configured to alternately send two-tone waveforms, with the same parameters but with different amounts of pre/post zero padding, to the oscilloscope for time alignment verification. The zero padding lengths were chosen so that the six waveforms all fit the chosen oscilloscope window of 36~\textmu s for each of the two channels. The verification waveforms were captured by the oscilloscope for the six nodes and saved by the GNU Radio software for post processing.

The SNR between all antenna pairs was set to 33$\pm$5~dB. A pulse-per-second (PPS) initial coarse alignment was used once at the start of each experiment to ensure that the transmitted time synchronization pulses arrived within the sampling window of the receiver window as described in \cite{merlo2022wireless}; this alignment provided only tens of nanoseconds of accuracy, and could be delivered either by one of the nodes in the array or by an external source such as GNSS. The SDRs were connected to a 10 MHz external frequency reference in node 0 and daisy chained to the other five nodes in the array (Fig.\ \ref{schematic}), providing frequency syntonization as described earlier; the approach may also be combined with wireless syntonization. Time division multiple access (TDMA) was used to sequence the time transfer between the nodes for the time synchronization; note that this approach requires only coarse time alignment (e.g., PPS) for scheduling. Two-way time transfer was used to estimate the time offsets between the local clocks via \eqref{delay_estimate} for each connection on a given node. Initially the network was fully connected, such that each node had an estimate of the offsets with respect to all other nodes. Subsequent experiments implemented randomly dropped connections, such that each node saw only a subset of connections. Here a specified number of connections was dropped at random such that the overall network connectivity was progressively lower, with a random graph produced during each iteration. The average consensus algorithm of Section II was implemented on each node to locally estimate the average offset, over time obtaining time synchronization in the full system.

\begin{figure}[t!]
\centering
\includegraphics[width=0.75\columnwidth]{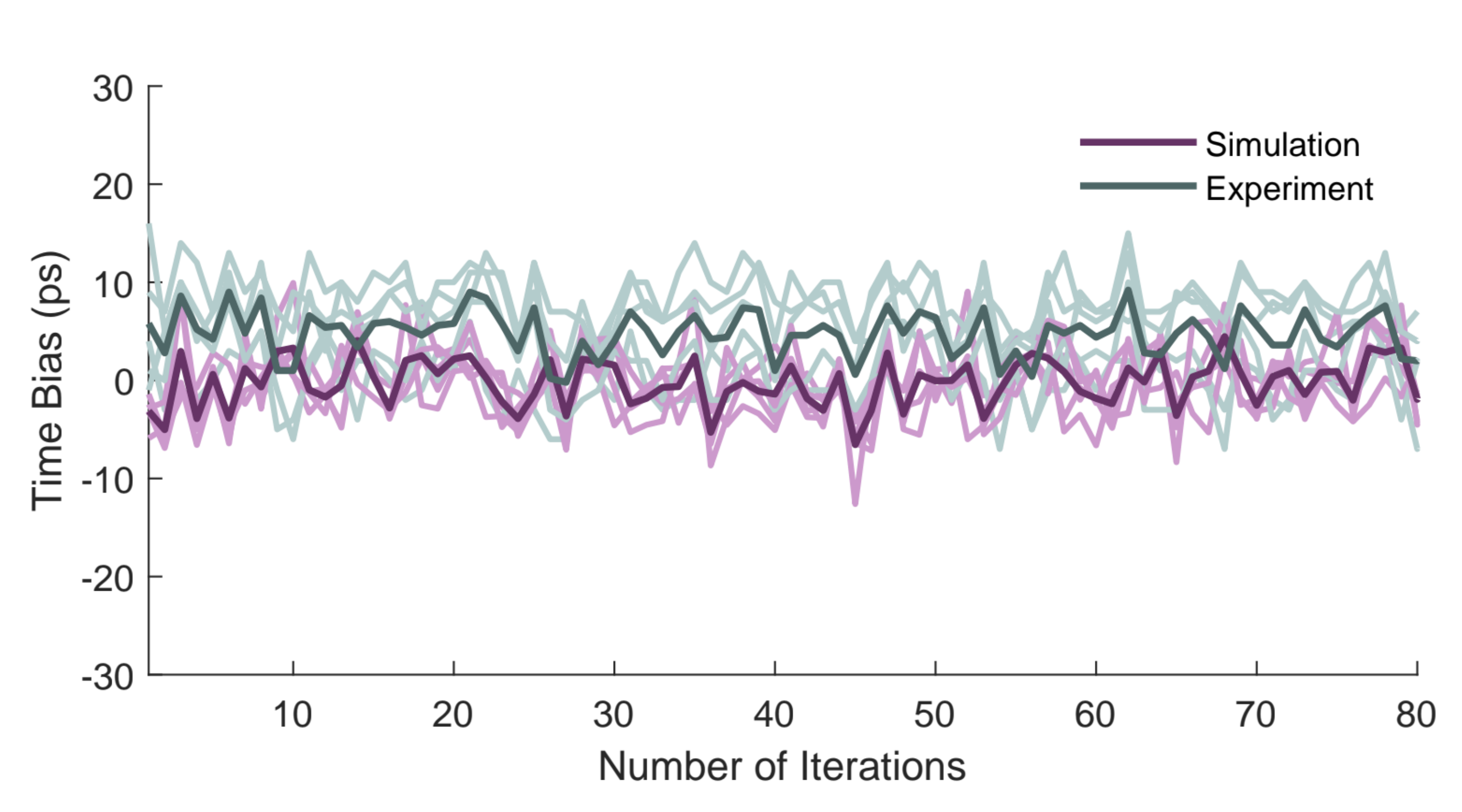}
\caption{Average (solid lines) and relative (shaded lines) biases for a six-node antenna array system. The decentralized time synchronization algorithm was applied to achieve time synchronization for the fully connected network.}
\label{average-bias_exp}
\end{figure}

The first simulation and experimental evaluation involved a fully connected six-node antenna array system. In this topology, the distributed array achieved average consensus after two iterations (two synchronization epochs), resulting in zero-mean bias (simulation) and under 10 ps average bias (experiment) and an average standard deviation under 3~ps. Fig.\ \ref{average-bias_exp} shows the relative and average time biases between the nodes. The residual biases observed at the oscilloscope are believed to be associated with propagation delay imbalances between the synchronization channel and the measurement plane. An assumption that is made is that the lengths of cabling between all channels after the power splitters are identical in the calibration channel, and in the measurement channel; any mismatches in these propagation delay due to cable lengths, or interconnects would appear as a residual bias.  Additionally, the different paths on the 3:1 power combiners connected to the oscilloscope may also have minor propagation delay mismatches. A time bias on the order of 5--10 ps is consistent with a path length mismatch on the order of 1 mm in PTFE ($\epsilon_r \approx 2.05$), which is reasonable given the number of cables and interconnects used. These biases were computed by finding the differences in time delay between the pulse received from node 0 and those received from the other nodes. 

\begin{figure}[t!]
\centering
\includegraphics[width=0.76\columnwidth]{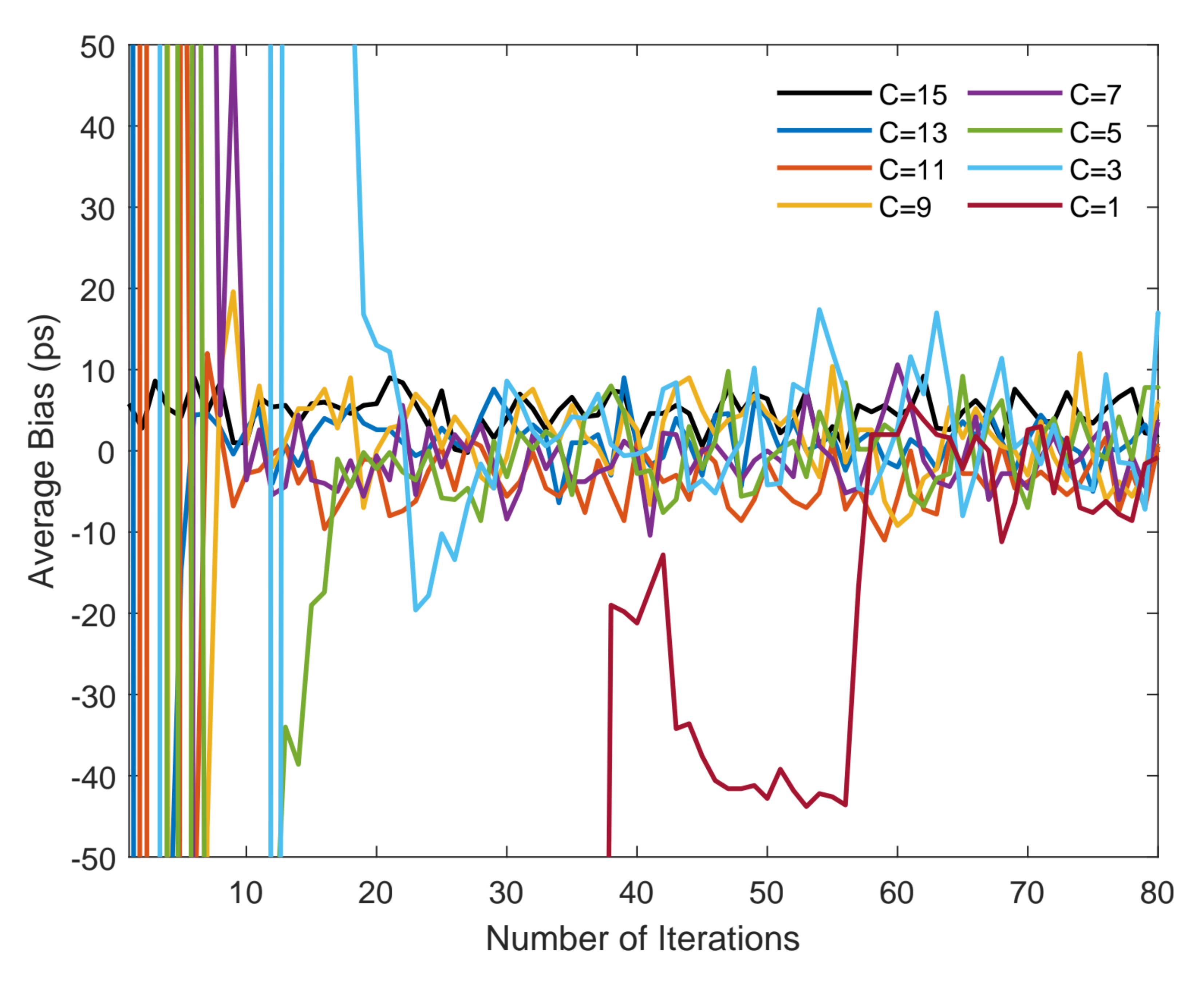}
\caption{Average bias for a six-node antenna array system. The decentralized time synchronization algorithm was applied to achieve time synchronization for the six nodes in the system with diverse connectivity. The number $C$ of connections was decreased by two in each experimental run. In each experiment, $C$ was fixed but in each iteration the dropped connection was chosen randomly.}
\label{average-bias}
\end{figure}

\begin{figure}[t!]
\centering
\includegraphics[width=0.75\columnwidth]{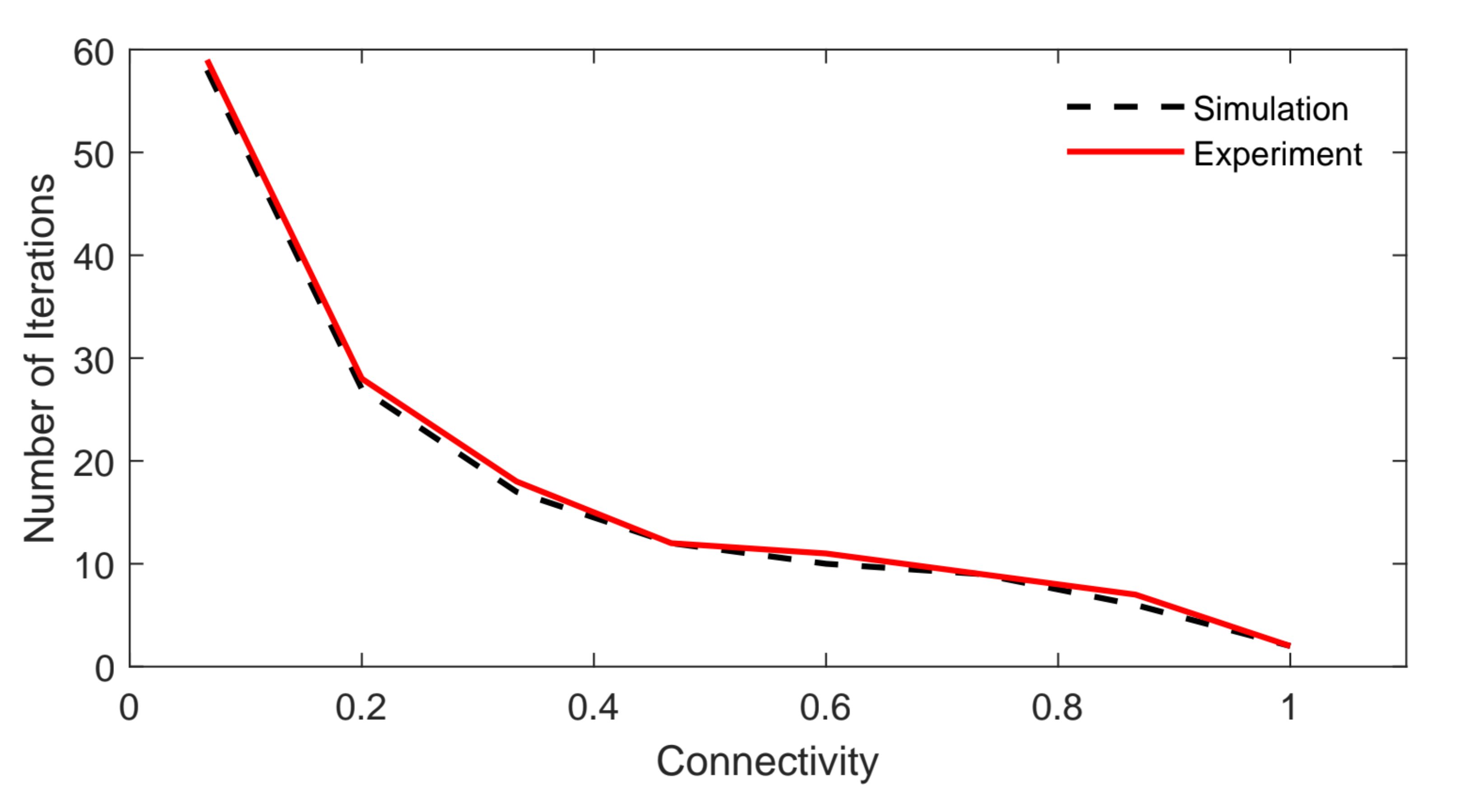}
\caption{Number of iterations required to achieve convergence at each connectivity (the number of connected nodes within the system divided by the total number of possible connections). The simulation results represent the average number of iterations from ten measurements.}
\label{iteration_vs_C}
\end{figure}

During the experimental evaluation, the received waveforms were captured, saved, and post processed using the same Python code that was used to process the simulation results. Post processing consisted of discrete matched filtering of the received signal with an ideal transmitted two-tone signal, detection of the matched filter output peaks along with the two adjacent points for each of the received signals, and refinement of the peaks by application of the quadratic-least-square (QLS) interpolation method. The peak refinement process is detailed in \cite{merlo2022wireless}. Note that time verification started after the first iteration because the time synchronization pulses and the verification pulses were sent during alternate iterations; the synchronization pulses were sent in first iteration for time alignment purposes, the verification pulses were sent in the second iteration, and so on. The sub-3 ps average standard deviation of the time offsets between the nodes was computed over 50 samples after 50 iterations. As was evaluated in simulation, the decentralized time alignment algorithm used in this work is unbiased. The experimental result shows an average bias of less than five picoseconds, which is due to systematic error as discussed previously. To evaluate system robustness to network connection diversity, the connectivity of the six-node system was varied by adjusting the weighting matrix in the software, synthesizing dropped connections between nodes. The relative time offsets between the nodes were computed and the average biases were plotted versus the iteration number. The baseline for this experiment was a fully connected network evaluated after 80 iterations. A fully connected six node system has $C=15$. Internode connections were dropped randomly during each iteration; first, one connection was dropped randomly ($C=14$) at each iteration for an 80 iteration run, then two connections were dropped ($C=13$), and so on. The number of iterations per experiment was held constant. The time delays of all the nodes were measured by the oscilloscope and estimated. Fig.\ \ref{average-bias} shows the average biases between the nodes for various values of $C$. The results from this experiment demonstrate the ability of the system to achieve and maintain time synchronization to within picoseconds under diverse connectivity conditions when the system is frequency syntonized. The effect of changing connectivity on system performance is further evaluated in Fig.\ \ref{iteration_vs_C}. The results agree with the simulation result and show that as connectivity decreases, the time required to achieve convergence increases.


\begin{figure}[t!]
\centering
\includegraphics[width=0.75\columnwidth]{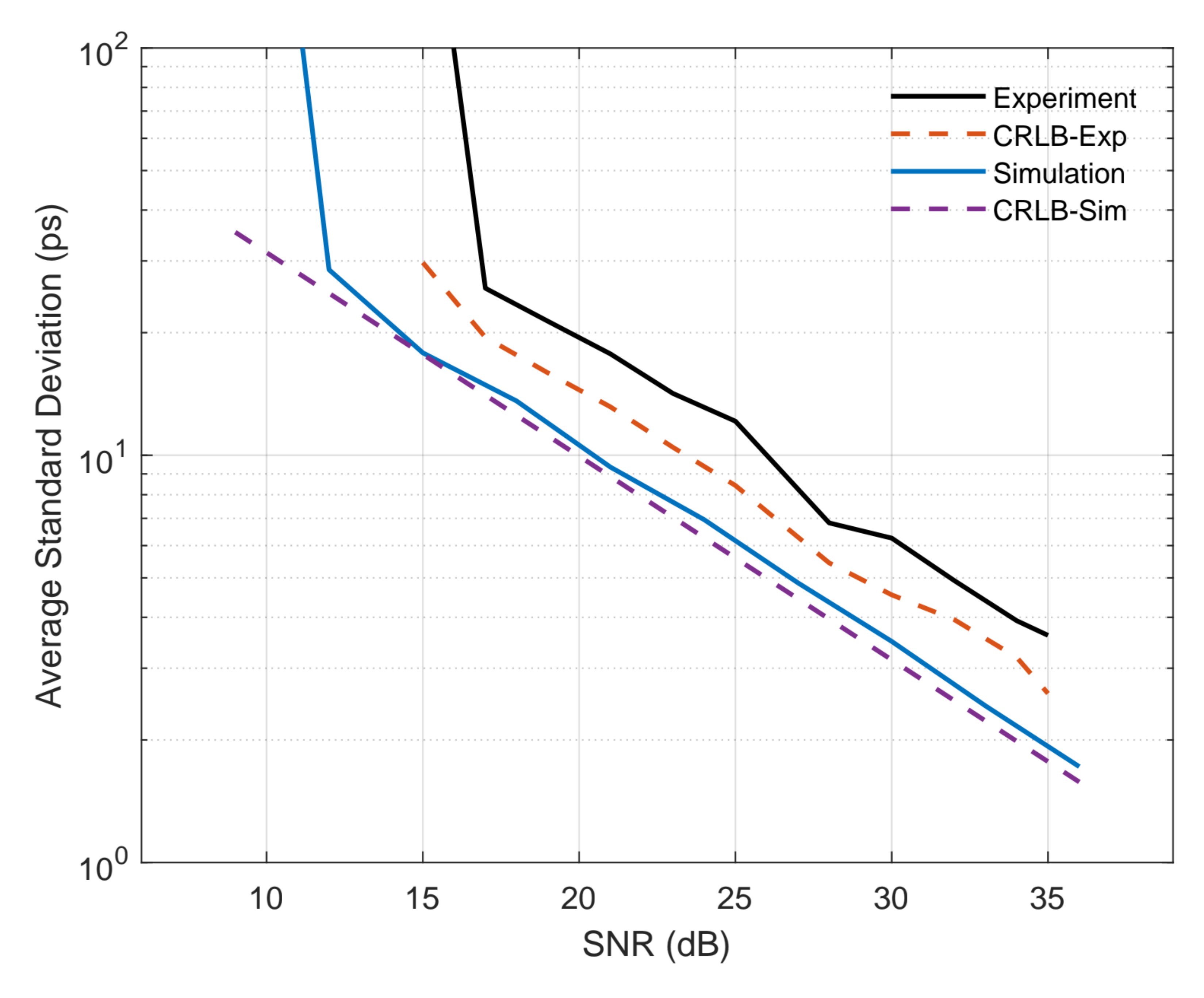}
\caption{The experimental result shows the average standard deviation in picoseconds for a six-node distributed array system evaluated over a range of 15--35 dB SNR. The CRLB values for all the internode links were averaged and the average CRLB is plotted for comparison as CRLB-Exp. In this experiment, the SNRs values were varied among the channels in the network. The simulation result shows the average standard deviation in picoseconds for a six-node distributed array system evaluated over a range of 9--36 dB SNR. The CRLB is plotted for comparison as CRLB-Sim. In this simulation, the SNR value was uniform over all channels in the network.}
\label{exp_SNR_std}
\end{figure}

Note that in Fig.\ \ref{average-bias} the average bias between the nodes for the case $C=1$ is still converging near iteration 50 (the first sample used in computing the standard deviation) and results in lower precision. The second reason for decreased precision is SNR variation among the links; this nonuniformity leads to increased error in the time information cycled between the nodes with links that have lower SNR values when the connectivity is low. Recall that as SNR decreases so does the precision; this explains the relationship between the link SNR and precision. However, in Fig.\ \ref{exp_SNR_std} the internode link SNR values are kept uniform in each run. To study the effect of nonuniform link SNR on the precision of the decentralized time alignment method, the method was applied to a fully connected network with nonuniform link SNR values, and the internode time offsets were measured. The transmitted and received gains for the nodes in the array were changed in each run so that the average estimated SNR varied from 15 to 35 dB. The time offset variances were computed over 100 iterations, and the average standard deviation at each average SNR value was computed using \eqref{CRLB2} and plotted in Fig.\ \ref{exp_SNR_std}. In contrast to the simulation of Section V, the nonuniform internode link SNR values do not allow direct comparison to the CRLB. Instead, individual link CRLB values were computed from the estimated SNR values, and the average CRLB was plotted for comparison. It is clear that the method precision follows the theoretical limit closely and increases along with the average link SNR. The range 15 to 35 dB was included because at average SNR of 15 dB at least one link had an SNR of 9 dB or less, which resulted in high average standard deviation due to the ambiguity of the two-tone signal (i.e., the estimated peak of the received signal does not represent the true peak).  

\section{Conclusion}

In this paper, a high accuracy wireless decentralized time synchronization method for a six-node distributed antenna array system was demonstrated and evaluated with dynamic network connectivity. The approach, which was implemented based on the average consensus and two-way time transfer methods, demonstrated an accuracy of less than 13 picoseconds and less than three picoseconds standard deviation (precision) along with unbiased time synchronization. The method showed robustness to network connection diversity.

\bibliographystyle{IEEEtran}

\bibliography{IEEEabrv,DTSDC}

\end{document}